\documentclass[aps,prd,reprint,superscriptaddress]{revtex4-1}
\usepackage[dvipsnames]{xcolor}
\usepackage[normalem]{ulem}
\usepackage{bm}% needed if you want to
\usepackage[utf8]{inputenc}
\usepackage{pgfplots}
\usepackage{qtree}
\usepackage{tikz}
\usepgfplotslibrary{groupplots}
\pgfplotsset{compat=1.14}
\usepackage{capt-of}
\usepackage{graphicx}% Include figure files
\usepackage{amssymb}
\usepackage{amsthm}
\usepackage{mathrsfs}
\usepackage{amsmath}
\usepackage{hyperref}
\usepackage{lineno}

\definecolor{pink}{RGB}{255, 20, 147}

\definecolor{byzantine}{rgb}{0.74, 0.2, 0.64}

\definecolor{amber}{rgb}{1.0, 0.75, 0.0}

\definecolor{amethyst}{rgb}{0.6, 0.4, 0.8}

\definecolor{blue-violet}{rgb}{0.54, 0.17, 0.89}

\definecolor{carminered}{rgb}{1.0, 0.0, 0.22}

\definecolor{Stef}{RGB}{102, 22, 97}

\definecolor{bittersweet}{rgb}{1.0, 0.44, 0.37}

\definecolor{amber}{rgb}{1.0, 0.75, 0.0}

\definecolor{ao(english)}{rgb}{0.0, 0.5, 0.0}

\definecolor{comment}{RGB}{166, 38, 164}

\begin{document} 
\title{The Timing System of LIGO Discoveries}

% These commands specify the author list and affiliations
\author{Andrew G. Sullivan}
\affiliation{Department of Physics, Columbia University in the City of New York, 550 W 120th St., New York, NY 10027, USA}
\author{Yasmeen Asali}
\affiliation{Columbia Astrophysics Laboratory, Columbia University in the City of New York, 550 W 120th St., New York, NY 10027, USA}
\author{Zsuzsanna M\'arka}
\affiliation{Columbia Astrophysics Laboratory, Columbia University in the City of New York, 550 W 120th St., New York, NY 10027, USA}
\author{Daniel Sigg}
\affiliation{LIGO Hanford Observatory, Route 10, Mile marker 2, Hanford, Washington 99352, USA}
\author{Stefan Countryman}
\affiliation{Department of Physics, Columbia University in the City of New York, 550 W 120th St., New York, NY 10027, USA}
\author{Imre Bartos}
\affiliation{Department of Physics, University of Florida, PO Box 118440, Gainesville, FL 32611-8440, USA}
\author{Keita Kawabe}
\affiliation{LIGO Hanford Observatory, Route 10, Mile marker 2, Hanford, Washington 99352, USA}
\author{Marc D. Pirello}
\affiliation{LIGO Hanford Observatory, Route 10, Mile marker 2, Hanford, Washington 99352, USA}
\author{Michael Thomas}
\affiliation{LIGO Livingston Observatory, Livingston, LA 70754, USA}
\author{Thomas J. Shaffer} 
\affiliation{LIGO Hanford Observatory, Route 10, Mile marker 2, Hanford, Washington 99352, USA}
\author{Keith Thorne}
\affiliation{LIGO Livingston Observatory, Livingston, LA 70754, USA}
\author{Michael Laxen}
\affiliation{LIGO Livingston Observatory, Livingston, LA 70754, USA}
\author{Joseph Betzwieser}
\affiliation{LIGO Livingston Observatory, Livingston, LA 70754, USA}
\author{Kiwamu Izumi}
\affiliation{JAXA Institute of Space and Astronautical Science, Chuo-ku, Sagamihara City, Kanagawa 252-0222, Japan}

\author{Rolf Bork}
\affiliation{LIGO Laboratory, California Institute of Technology, Pasadena, CA 91125, USA}
\author{Alex Ivanov}
\affiliation{LIGO Laboratory, California Institute of Technology, Pasadena, CA 91125, USA}
\author{Dave Barker}
\affiliation{LIGO Hanford Observatory, Route 10, Mile marker 2, Hanford, Washington 99352, USA}
\author{Carl Adams}
\affiliation{LIGO Livingston Observatory, Livingston, LA 70754, USA}
\author{Filiberto Clara}
\affiliation{LIGO Hanford Observatory, Route 10, Mile marker 2, Hanford, Washington 99352, USA}
\author{Maxim Factourovich}
\affiliation{Department of Physics, Columbia University in the City of New York, 550 W 120th St., New York, NY 10027, USA}
\author{Szabolcs M\'arka}
\affiliation{Department of Physics, Columbia University in the City of New York, 550 W 120th St., New York, NY 10027, USA}

\begin{abstract}
LIGO's mission critical timing system has enabled gravitational wave and multi-messenger astrophysical discoveries as well as the rich science extracted. Achieving optimal detector sensitivity, detecting transient gravitational waves, and especially localizing gravitational wave sources, the underpinning of multi-messenger astrophysics, all require proper gravitational wave data time-stamping. Measurements of the relative arrival times of gravitational waves between different detectors allow for coherent gravitational wave detections, localization of gravitational wave sources, and the creation of skymaps. 
The carefully designed timing system achieves these goals by mitigating phase noise to avoid signal up-conversion and maximize gravitational wave detector sensitivity. 
The timing system also redundantly performs self-calibration and self-diagnostics in order to ensure reliable, extendable, and traceable time stamping.
In this paper, we describe and quantify the performance of these core systems during the latest O3 scientific run of LIGO, Virgo, and KAGRA. We present results of the diagnostic checks done to verify the time-stamping for individual gravitational wave events observed during O3 as well as the timing system performance for all of O3 in LIGO Livingston and LIGO Hanford. We find that, after 3 observing runs, the LIGO timing system continues to reliably meet mission requirements of timing precision below 1 $\mu$s with a significant safety margin. \\

\end{abstract}
\maketitle

% -------------------------------------------
\section{Introduction}
\label{sec:Introduction} 
% -------------------------------------------

The LIGO gravitational wave (GW) detectors have completed three GW observing runs since becoming operational in 2015 \cite{2010CQGra..27h4006H, 2015CQGra..32g4001L, 2016PhRvL.116m1103A, 2016PhRvD..93k2004M, 2019PhRvL.123w1107T, 2020PhRvD.102f2003B, 2020LRR....23....3A}. The network of LIGO, Virgo \cite{2015CQGra..32b4001A}, GEO600 \cite{2016CQGra..33g5009D, 2014CQGra..31v4002A}, and KAGRA \cite{2013PhRvD..88d3007A, 2021PTEP.2021eA101A} detectors has successfully observed GWs from a large number of astrophysical sources including binary black hole (BBH) and binary neutron star (BNS) inspirals \cite{PhysRevX.9.031040,2021PhRvX..11b1053A, 2021arXiv211103606T}. 

In order to make these discoveries and coordinate between the global GW detector and multi-messenger followup networks, both LIGO Hanford (LHO) and LIGO Livingston (LLO) require a precise mechanism to determine timing of its observed gravitational wave events. Not only must timing be maintained between the global detector network, but the 4 km size of the interferometers necessitates precise timing at all parts of the individual detector across large distances. Consequently, the LIGO timing system was developed to ensure precise time recording at all relevant parts of the detectors \cite{aLIGOtiming, aLIGOtimingdiagnostics}. The creation of the timing system was a collaboration between Columbia University, LHO, and the Caltech and is maintained by Columbia, LHO, and LLO. 

The operation of the timing system during and in the followup of an observing run is a mission-critical aspect of LIGO as reliable interferometer operation and astrophysical data analysis both directly depend on it. The timing system and its carefully selected components allow (a) phase noise to be minimized via its precise crystal oscillators, lessening the timing contribution to the detector noise level and consequently allowing LIGO's astrophysical reach, (b) the absolute timing of the data streams in the detector network to agree to high accuracy, making coincident and coherent observation using the GW detector network possible, (c) verification of data-stream timing across detectors so that a sufficiently numerous network of interferometric detectors can recover both the polarization and sky direction information for a detected event to high accuracy and precision, and (d) quick and trustworthy timing information on hand for multi-messenger searches for coincident detections of GWs and other astrophysical events, such as gamma ray bursts (GRBs) or supernovae \cite{2008CQGra..25k4051A,070201,2010JPhCS.243a2001M,PhysRevLett.119.161101}. 
The timing and timing distribution systems are essential for proper GW data collection and provide end-to-end witness signals for wholistic assurances on data timestamps. 

With three observing runs complete, the LIGO timing systems have produced a large sum of data with which one can assess their functionality and performance. In this paper, we discuss how the timing system and the timing diagnostic system have performed during the third GW observing run.
Other systems outside of the timing systems, such as the interferometer components, the photodetectors, the calibration, the filters, or the electronics, can on occasion introduce timing delays. Analyses of the effects of these other systems on the timing system require separate investigations, not covered by this study.

 We divide the text as follows. In Section \ref{sec:distribution_system}, we review the configuration of the LIGO timing system. In Section \ref{sec:timeverification}, we discuss the timing verification studies done to assess the accuracy of the system and present the results of these studies from O3. In Section \ref{sec:DiscussionandConc}, we discuss the results of the O3 timing verification studies and conclude.

% -------------------------------------------
\section{Advanced LIGO Timing System}
\label{sec:distribution_system}
% -------------------------------------------

\subsection{Timing Distribution System}
The timing distribution system provides synchronized timing (a) to an absolute time measure (UTC) so that results can be compared between LHO, LLO, other members of the global GW-detector-network, and non-GW detectors to enable multi-messenger astronomy; (b) to subsystems of a given detector, ensuring the reliable operation and control of the interferometer; and (c) with minimized data acquisition timing signal jitter. The timing distribution system thus creates the universal, precise, and traceable time-frame in which the components of the entire interferometer network are synchronized \cite{LIGO-T020036-x0, LIGO-T070218-x0}. The timing system is also redundantly cross checked to time sources of different Allan-variance \cite{1966IEEEP..54..221A} to protect against single technology failure.

The timing precision required for the LIGO detectors varies for different science goals and evolves with discoveries of new science. Consequently, a compromise was adopted, with an engineering specification requirement for intra-observatory systems of better than 1 $\mu$s \cite{aLIGOtiming} performance at the very worst case; however, to allow for future scientific advances, it was recommended that the timing system should achieve the best possible performance allowed by technology and other constraints available at the time of its design. The timing distribution system \cite{aLIGOtiming} has a hardware defined timing accuracy and overall clock synchronization precision that satisfies LIGO requirements (laboratory measurement indicates ``median of 14.3\,ns and a 99\% confidence interval of 6.2\,ns (with maximum of 24\,ns and minimum of 10.5\,ns)'' \cite{aLIGOtiming,T080083,T0900050} deviation between the timing reference input and trigger passed to data acquisition system (DAQ), assuming that the fiber pair used for the fiber delay measurement is symmetric in length). 

At each LIGO site, the primary synchronization reference is derived from the signals of multiple Global Positioning System (GPS) satellites \cite{_gps.gov}, as a root source of timing. A GPS backed clock at the corner station \cite{LIGO-D1002546-v3, LIGO-T1000659-v2} receives timing information from a GPS receiver \cite{trimGPSantenna} and distributes it to the system. A comparison flywheel of a Cs-III atomic clock \cite{csiiI4301B} is continuously monitored to identify GPS and other systemic transient problems. These atomic clocks run fully independently from the GPS system after their reset and initial precision calibration before an observing run \cite{LIGO-T1500423-v2}. Since even calibrated Cs clocks display a slight drift, these atomic clock comparisons are not used to recover absolute time. 

The timing distribution system is structured as a chain of modules beginning from the trunk module, extending to branch modules that fan out to various leaf modules. The trunk module receives GPS timing data from a receiver module as a 1PPS signal and synchronizes its internal oven-controlled crystal oscillator (OCXO) clock to the GPS time. The OCXO operates on a $2^N$ Hz base frequency where $N=26$ during O3. The trunk then passes the information to branch modules which contain a voltage-controlled crystal oscillator (VCXO) clock. The branch modules synchronize their internal clock to the received timing signal. Individual branch modules may fan out to as many as 16 other branch or leaf modules \cite{aLIGOtiming}. The leaf modules receive the timing signal from a branch module and synchronizes their internal clocks (also a VCXO clock) to it. The leaf module passes on timing information to the rest of the detector through the use of daughter boards. A timing converter on the leaf module sends timing pulses (as well as Inter-Range Instrumentation Group B (IRIG-B)\cite{T1000477,T1100574} and DuoTone signals \cite{E0900019,DuoToneCharacterization}) to the LIGO DAQ.
The Timing System thus allows for a plug-and-play dynamic topology, e.g.
\smallskip

\Tree [.Trunk [.B I
                    [.B D
                         D
                         I ]
                    C ]
               C
               I
               [.B D
                    D
                    I
                    C
                    C ]
               D ]

\begin{figure*}[htbp]
    \centering
    \includegraphics[width=\linewidth]{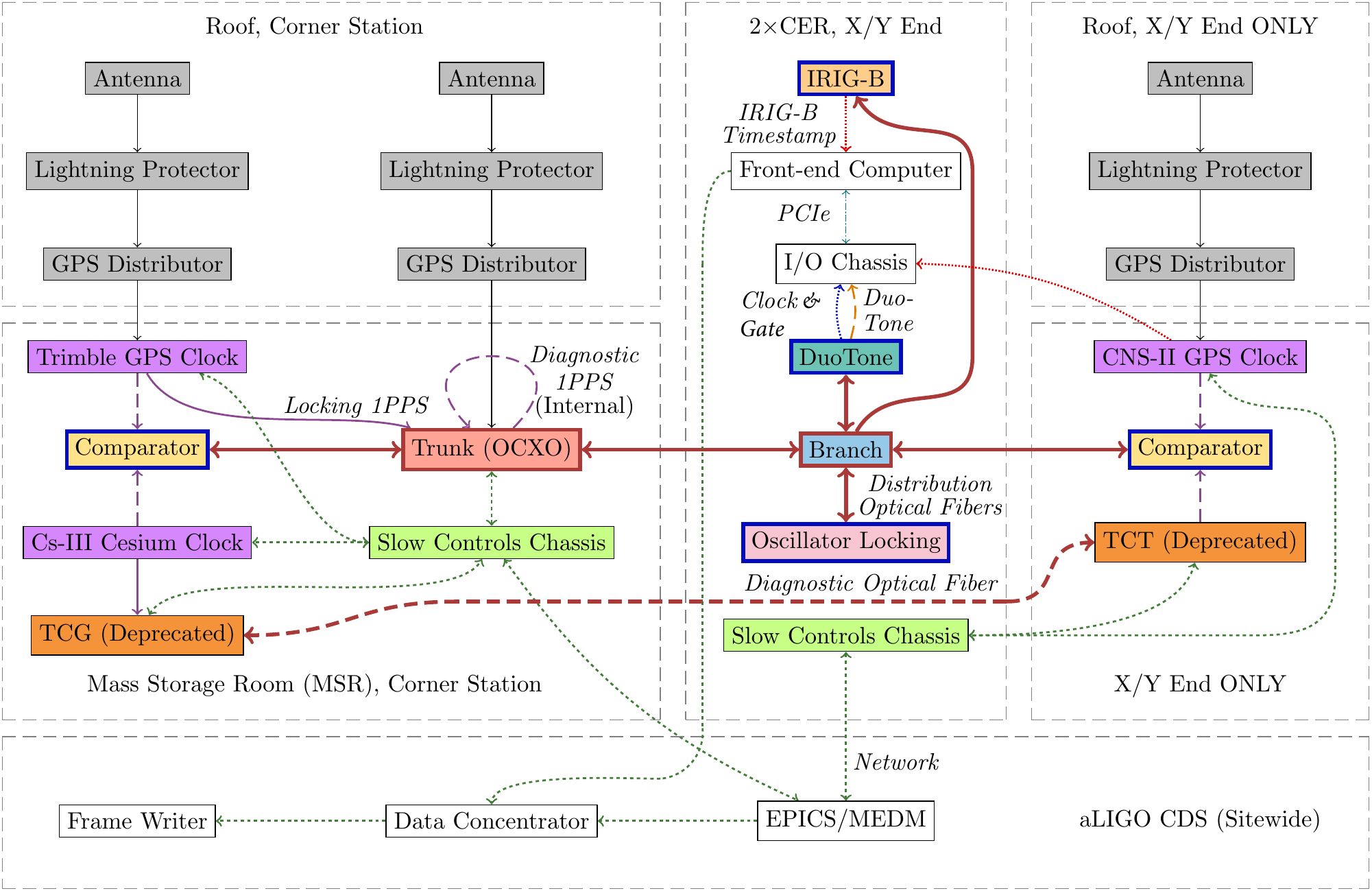}
     \caption[Timing System information flow]{
                Timing System information flow.
                The gray dashed panels represent different locations within the interferometer.
                Includes both \textit{distribution} and \textit{diagnostic} signals.
                A Trimble GPS clock passes 1PPS signal to the trunk, which synchronizes its OCXO clock to the signal.
                The trunk's built-in GPS clock provides a diagnostic 1PPS source as well as the GPS timestamp during normal operation.
                The trunk distributes the timing signal to leaf modules (via branch modules) through optical fibers. Leaf-DuoTone assemblies installed in I/O chassis produce a $2^{16}$ Hz clock signal (with $<1$ $\mu$s error) and a DuoTone signal (digitized at $2^{16}$ Hz and decimated down to  $2^{14}$ Hz) for diagnostics and calibration.}
    \label{fig:TimingSystem}
\end{figure*}

\smallskip
where the symbol meanings are as follows: \textit{B} is a \textit{Branch}, \textit{C} is a \textit{Comparator},  \textit{I} is a \textit{IRIG-B}, and \textit{D} is a \textit{DuoTone-Leaf} module.
Figure \ref{fig:TimingSystem} shows a diagram of the timing system, including the path of both the timing and diagnostic signals around the Trunk as well as at the branches and leaves. 
 
Every board in the timing distribution chain was meticulously tested by different teams and in different environments, including end-to-end laboratory tests using long fibers. It is essential that the accuracy of the distributed timing signals is closely monitored to verify the accuracy of the gravitational wave data stream and ensure that it can be correlated with other data streams. 

The LIGO detectors (and most other observatories) rely on GPS to ensure long-term UTC synchronized timing. Consequently, there is a relatively small additional correction to the upper limit on the inter-site timing accuracy, as the GPS system is only guaranteed for hundreds of ns. In practice, GPS does much better most of the time and, even if the most conservative estimate is used, the absolute total timing discrepancy between detectors would not affect the utility of current gravitational wave observations \cite{aLIGOtiming}. The GPS system is closely followed by multiple agencies and usually performs at the $\sim50$ ns level \citep{GPSGOVstandard,FAAstandard}. In principle, LIGO's timing system also allows for the detection of GPS system problems as relevant data is recorded.  

\subsection{Timing Diagnostic System}
The mission critical nature of the LIGO timing distribution system requires the system's status be closely monitored, periodically tested in-depth, and traceable. The timing diagnostic system has the following major components:

\noindent
1. \textit{Self Diagnostics ---} The timing system has extensive and continuously recorded self-diagnostic information implemented in hardware. Every element in the timing distribution chain sends back detailed diagnostic information (as shown in Figure \ref{fig:TimingSystem}) which is archived and also available through Motif Editor and Display Manager (MEDM) screens in the LIGO Control Room. The collected information includes diagnostics such as fiber delays and control voltage for the different flywheels within the system. Additionally, information about the structure, topology, and configuration of the system as a whole is monitored and archived along with the connection and up/down status of each node. MEDM screens in the LIGO control rooms show if errors are present in real time. This self-diagnostic data is available for further analysis and characterization. 

\noindent
2. \textit{Multiple, cross-verified time sources ---} Many test-points of the timing distribution system are compared to independent time sources of different manufacturer, firmware, type, and location. These include additional GPS synchronized time sources connected to independent GPS antennae and a Cs-III atomic clock at each site. The workhorses of the timing diagnostic system are the timing comparators \cite{T1200331}. They compare 1PPS signals from many devices to the 1PPS signal from the timing distribution chain with high precision. The comparator timing chassis reports the measured time differences in every second that is recorded along with the other diagnostic information mentioned above.

\noindent
3. \textit{Timing diagnostics ---} There are several additional hardware-generated synchronized timing witness channels available for timing diagnostics, such as the so-called DuoTone signals produced on each Leaf-DuoTone \cite{T1000670} assembly within each LIGO analogue-to-digital converter (ADC) chassis. DuoTone signals are digitized and recorded along with the GW data channel as a witness of precise timing. The phase of the DuoTone signal allows sub-microsecond accurate determination of the digital datastream's shift from perfect agreement with the UTC time.

An additional witness is the IRIG-B \cite{T1000477,T1100574} signal. The timing distribution system itself includes IRIG-B interfaces, which produce IRIG-B signals used for communicating with devices that require IRIG-B input, including some computer interfaces. The independent GPS sources at the End Stations also have IRIG-B output capability for added redundancy and accountability. The IRIG-B signal has a phase that allows time verification on the ms level while its full timecode allows the determination of absolute YEAR:MONTH:DAY-HOUR:MINUTE:SECOND unambiguously. Independent IRIG-B signals at the End Stations are digitized and used to verify the timestamp provided by the timing distribution system.

The timing diagnostics hardware records multiple robust diagnostics and witness signals covering all possible time-shifts. It also provides independent verification of the accuracy and validity of the recorded timestamp for a particular piece of data.

\section{Timing Verification Studies}
\label{sec:timeverification}

We now discuss the various timing verification studies conducted during the observing runs of LIGO and present the results of these studies from O3. 
\\

\noindent
\textit{DuoTone Delay Checks---} When precision measurement of sub-sample timing in a digital system is required, sinusoid-shaped witness signals are well suited due to their low impact. To measure longer times than the period of a sinusoid, multiple sinusoids of slightly different frequencies can be added, extending the range of measurement. The Leaf-DuoTone assembly provides a two tone signal with frequencies 960 Hz and 961 Hz \cite{aLIGOtiming}. This choice bridges the gap between the IRIG-B's subsecond accuracy and the system's reach into the ns range, giving us the ability to characterize delays over 16 orders of magnitudes, from ns to years. 
The use of a sinusoid timing signal composed of only two high frequency lines mitigates potential DAQ crosstalk within the
frequency range of expected astrophysical binaries. Additionally, these frequencies are chosen around 960 Hz to coincide with the higher harmonics of the 60 Hz power-line frequency, which already contaminates LIGO's bandwidth. 

\begin{figure}[htbp]
    \centering
    \includegraphics[width=\columnwidth]{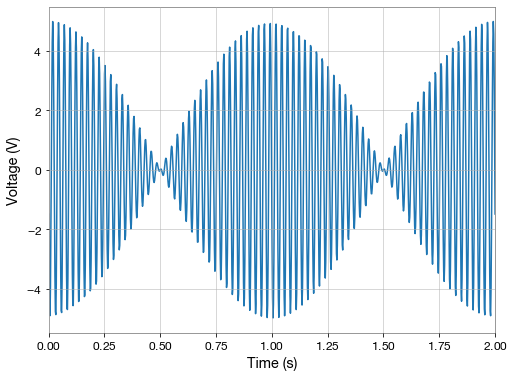}
    \includegraphics[width=\columnwidth]{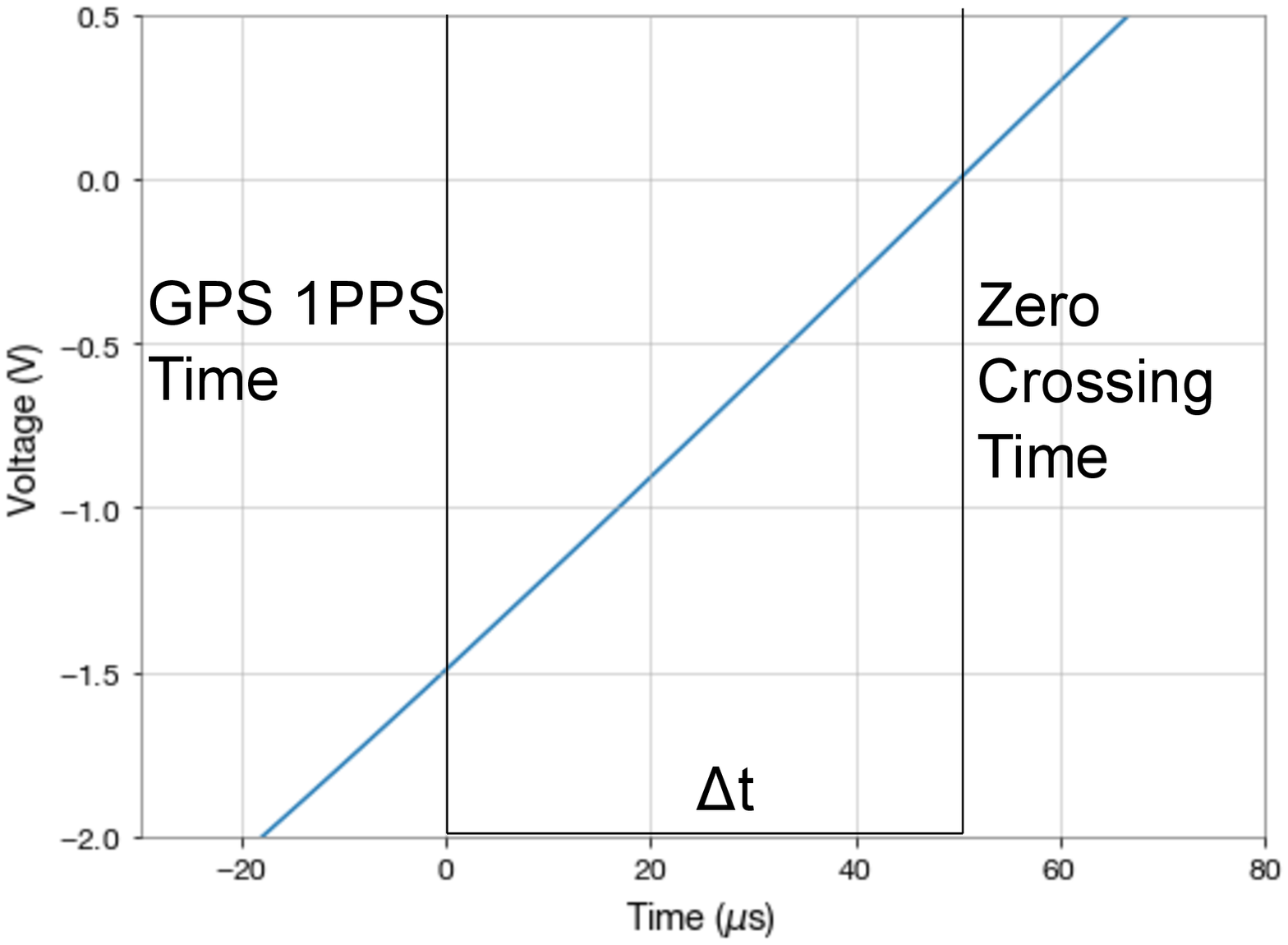}
    \caption{Plots of a simulated DuoTone diagnostic signal that would be created by the Leaf-DuoTone board including a characteristic offset $\Delta t$. (Top) Two seconds of the DuoTone signal. (Bottom) A closeup around the zero crossing time of the DuoTone signal, with the time shifted so that $t=0$ is at the time of the GPS 1PPS signal rising edge. The time of the GPS 1PPS rising edge and the zero-crossing time are marked on the plot.}
    \label{fig:duotoneex}
\end{figure}

The coincident zero crossing time of the DuoTone signals are hardware synchronized to the GPS 1PPS rising edge. The use of only two frequencies allows for unambiguous delay measurement at the second scale, providing precision enhancement to IRIG-B. Of course, TriTone or QuadTone signals can go beyond the second scale limitation of DuoTone without compromising precision significantly. The digitized DuoTone diagnostic signal is compared with GPS timing. Deviations between the DuoTone zero crossing time and GPS 1PPS rising edge may be observed with ns precision. The DuoTone diagnostic signal $Y$ may be represented as follows 
\begin{subequations}
    \begin{equation}
        \label{eq:Y1}
        Y_1=A\sin{(2\pi\times960(t-\Delta t))}
    \end{equation}
    \begin{equation}
        \label{eq:Y2}
         Y_2=A\sin{(2\pi\times961(t-\Delta t))}
    \end{equation}
    \begin{equation}
        \label{eq:Diag}
        Y=Y_1+Y_2+\Delta A
    \end{equation}
\end{subequations}
where $A$ is the amplitude of the DuoTone signals with a value of $2.5$ V, $\Delta A$ is centered at $0$ V, and $\Delta t$ is the phase difference between the DuoTone signal and the GPS 1PPS rising edge \cite{DuoToneCharacterization}. The DuoTone signal naturally obtains a delay from the GPS time. The expected delay, as measured \cite{DuoToneCharacterization} on a set of characteristic board combinations, is $50.25$ $\mu$s (6.70 $\mu$s from the designed-in delay between the DuoTone signal and the timing branch and $43.55$ $\mu$s from the 65536 Hz to 16384 Hz decimation filter at 960 Hz \cite{LIGOO2duotone}). The residual delay after subtracting $50.25$ $\mu$s from $\Delta t$ at each time segment is the uncertainty in the timing measurement. Figure \ref{fig:duotoneex} shows a plot of Equation \ref{eq:Diag}. As Figure \ref{fig:duotoneex} shows, the zero crossing time in the center of a pulse occurs once per second, so the time difference $\Delta t$ between the zero crossing time and the time of the GPS 1PPS rising edge can be measured for each second.

\begin{figure}[htbp]
    \centering
    \includegraphics[width=\columnwidth]{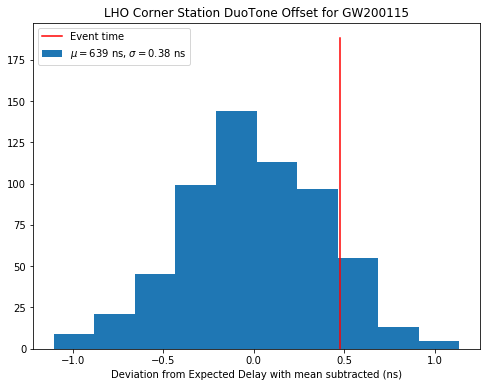}
    \caption{DuoTone delay histogram for the Corner Station at LHO in the $\pm5$ minutes surrounding GW200115 with mean delay $\mu$ subtracted. The standard deviation $\sigma$ is also reported in the legend. The residual time at the GW event time is denoted by the position of the red line.}
    \label{fig:duotone}
\end{figure}

DuoTone delay checks are always performed in low-latency for the interval of 5 minutes before and after a GW event to ensure that the GPS time-stamping of events is highly accurate at the detectors. We perform DuoTone delay checks for all GW event candidates, observing the residual time difference between the GPS 1PPS rising edge and DuoTone zero crossing time. Checks are performed at the Corner Stations of both LHO and LLO at a point close to a digitizer where the most critical signals for gravitational-wave detection are recorded. Checks are also performed at the X and Y End Stations of both LHO and LLO at points close to digitizers that record mission critical signals for the calibration of the detectors ~\citep{2016RScI...87k4503K}. For each check, we create a histogram for the observed residual delay over the course of the $\pm5$ minutes of the GW event. Figure \ref{fig:duotone} shows a DuoTone histogram at the Corner Station of LHO during the $\pm5$ minutes of GW200115, one of the Neutron Star-Black Hole merger events observed in O3B \cite{Abbott_2021}. Plots like this are also made for the DuoTone delay at the X and Y End Stations of both detectors. As the plot demonstrates, the residual times observed for GW200115 were well below the 1 $\mu$s threshold at LHO and thus GW200115 is properly time-stamped at that detector. 

\begin{figure*}[htbp]
    \centering
    \includegraphics[width=2\columnwidth]{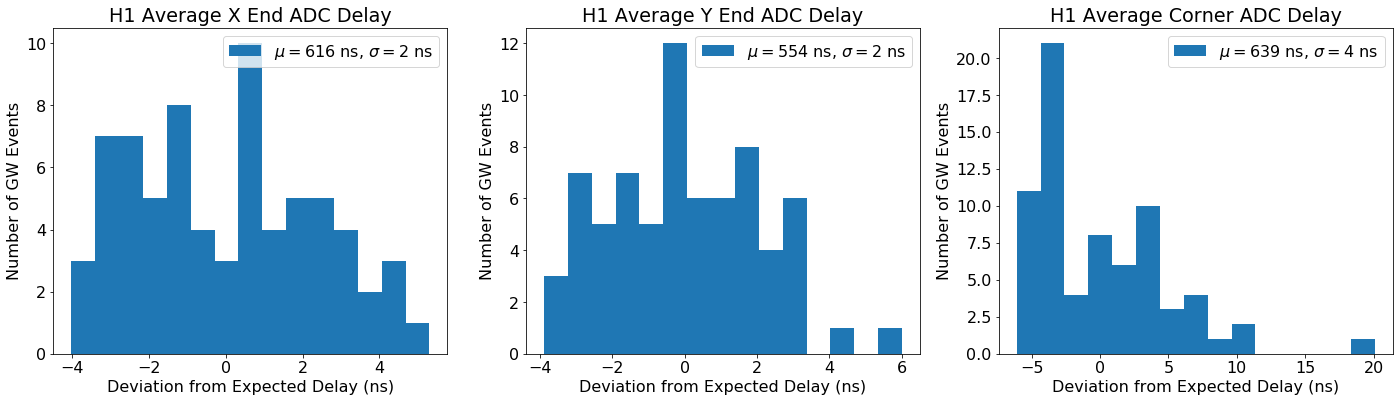}
    \includegraphics[width=2\columnwidth]{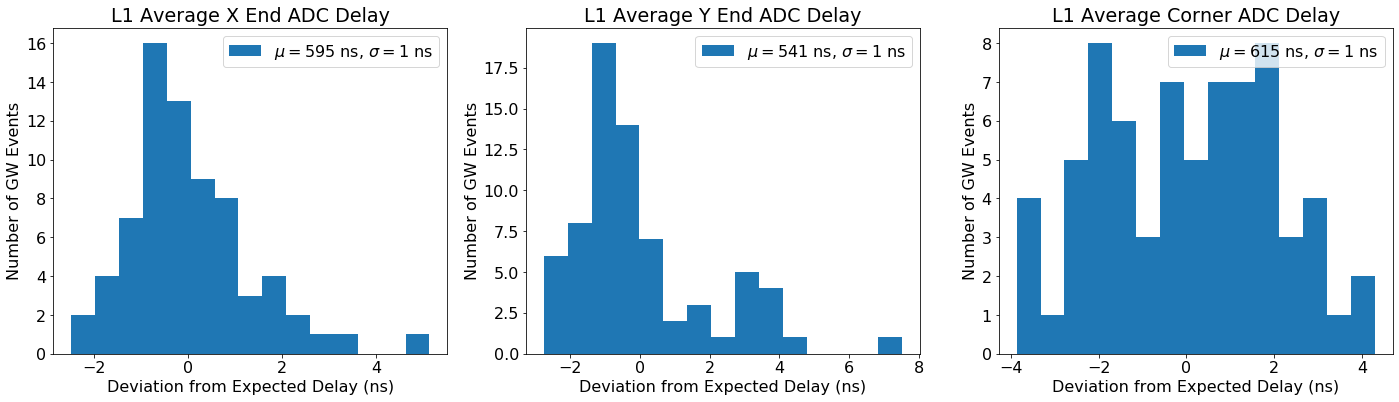}
    \caption{The average ADC DuoTone time delay $\pm 5$ minutes around all GW event observed during O3 in H1 (Top) and L1 (Bottom). Each bin contains the number of GW events with a certain average time delay.  The mean delay across all events $\mu$ has been removed. The legend of each subplot shows $\mu$ along with its standard deviation $\sigma$. Each subplot shows the delay in a different location denoted by the plot title.}
    \label{fig:duotoneO3}
\end{figure*}

\begin{figure*}[htbp]
    \centering
    \includegraphics[width=2\columnwidth]{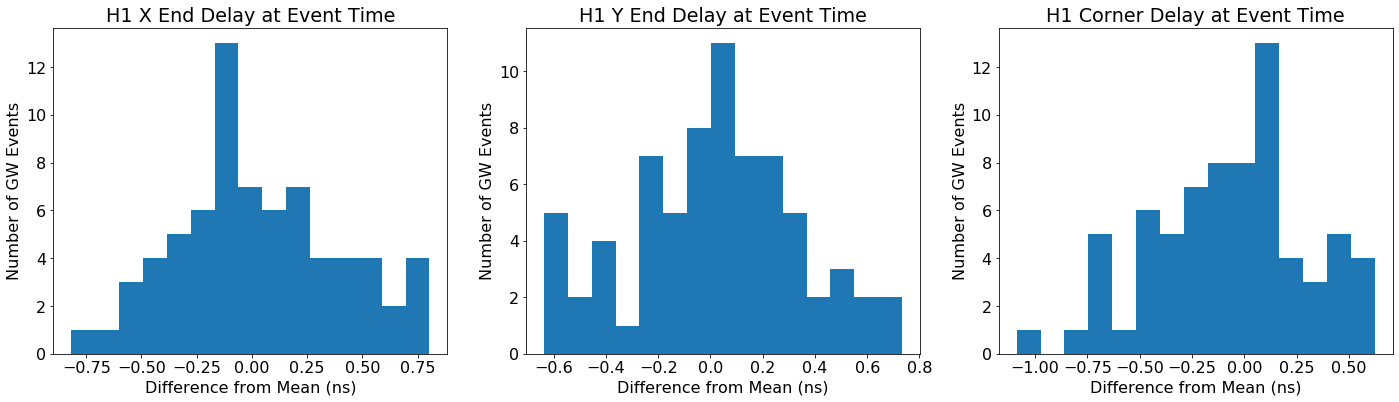}
    \includegraphics[width=2\columnwidth]{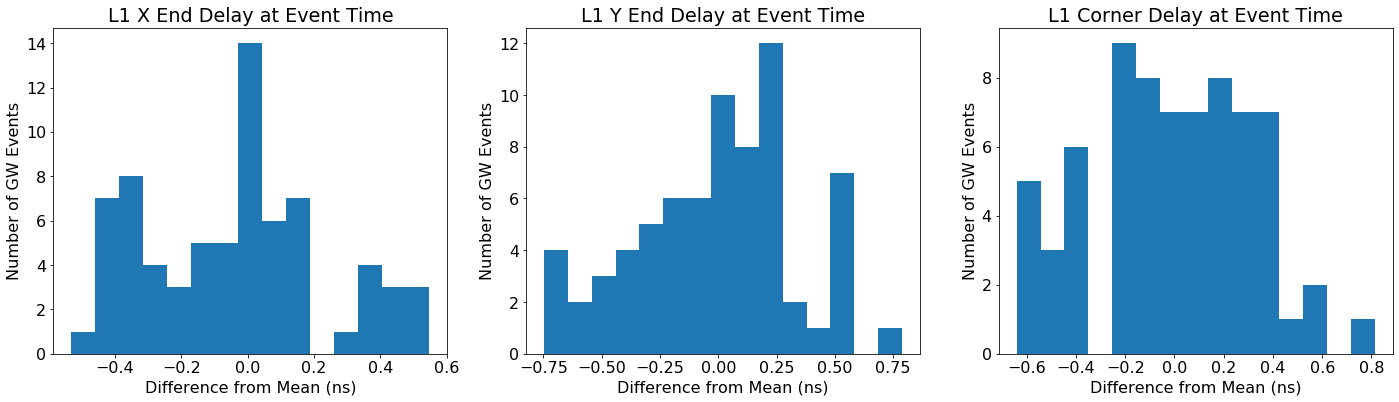}
    \caption{The differences between the DuoTone delay at the GW event times and the average DuoTone time delay in the $\pm 5$ minutes around the GW events observed during O3 in H1 (Top) and L1 (Bottom). Each subplot shows the delay difference at a different detector location.}
    \label{fig:duotoneO32}
\end{figure*}

Figure \ref{fig:duotoneO3} shows histograms of the average time delay over the $\pm5$ minutes around every GW event in O3. We also plot histograms of the difference between the DuoTone delay at the event time and the average delay around that event in Figure \ref{fig:duotoneO32}. The DuoTone checks returned residual time variations below 1 $\mu$s in the ADC channels for all GW events of O3 at both LLO and LHO. The average delay of each event never differed from the overall mean delay by more than $\pm20$ ns at any location. Furthermore, the difference between the delay at the event time and average event delay never exceeded $\pm1.1$ ns. Consequently, the DuoTone checks demonstrate timing precision within the designed threshold \cite{aLIGOtiming} for all GW events in O3. These checks were performed for GW event candidates in previous observing runs as well \cite{LIGO-T1600161,LIGO-T1500516,LIGO-T1700484-v1,LIGO-T1700483-v1,LIGO-T1700482-v1,LIGO-T1700438-v1,LIGO-T1700437-v1,LIGO-T1700384-v1,LIGO-T1700039-v2,LIGO-T1700306-v1,LIGO-T1700305-v1,LIGO-T1700304-v1,LIGO-T1700303-v1,LIGO-T1700302-v2,LIGO-T1700077-v1,LIGO-T1700041-v1,LIGO-T1600161-v1,LIGO-T1500516-v2}.

\noindent
\textit{IRIG-B Checks--} IRIG-B timing verification checks function by decoding the IRIG-B interface's time-stamping (converted from the internal LIGO timing system signal encoding) and comparing it with GPS time to ensure that at least on the order of ms the LIGO timing system remains aligned with GPS time. Because the IRIG-B signal has ms precision due to a measurable phase, it may be used as a corroborative check in addition to the DuoTone checks to ensure there are no timing issues around GW events. 

The IRIG-B encoded signal produces high and low voltage pulses. The length of these pulses determines whether a given pulse corresponds to a value of 0, 1, or a control character. A short high voltage pulse corresponds to a value of 0, a long high voltage pulse corresponds to a value of 1, and a very long pulse corresponds to a control. The control pulses regulate the phase of the IRIG-B signal. Additionally, for each unit of time (seconds, minutes, etc.) of the timestamp, a high voltage pulse corresponds to the number of that unit. To obtain the time encoded manually, one must add the time values of each long pulse for each unit of time. 

For each GW event observed by LIGO, the IRIG-B time at the End Station of each interferometer is checked against the GPS time in the 30 second window surrounding the time of the GW event. For all GW events in O3, the IRIG-B output time matched the GPS time. We observed no issues with the IRIG-B timestamping at either LLO or LHO for any of the GW candidates observed in O3. We note, however, there was a  persistent firmware problem in the LHO Y End Station CNS II clock that affected the IRIG-B timing output. When the IRIG-B time changed from 23:59:59 to 0:00:00, the IRIG-B would output the incorrect time at 0:00:00, but would output the correct time at all other times of the day. This issue was identified in LLO and LHO prior to O3; however, it was only fixed in LLO \cite{T1700065-v2} and thus remained in the LHO Y End Station. After O3 was completed, the issue was fixed in LHO as well. As this was an issue with the diagnostic CNS II clock, not the timing system itself and none of the GW events coincided with the times of this issue, the timing system's performance and all GW event data remain unaffected.

\noindent
\textit{Slow Channel Checks--} Throughout an observing run, the 1PPS signals from the timing distribution are continuously compared with independent GPS clocks at various locations in the LIGO detectors. This ensures that the clocks at the various stations are properly synchronized and that timing of data and systems at different parts of the LIGO detectors properly correspond to GPS time. The three most important clocks considered in these checks are the GPS backed NTP server clock \cite{NTP} found in the computer room of the Corner Station of the LIGO detectors and the CNS II GPS clocks \cite{CNSii} housed at the two End Stations of each LIGO detector. The time difference between these clocks and the LIGO timing distribution's 1PPS signal is recorded throughout the observing runs. Assessments of the slow channel timing performance have been conducted for O1 and O2 \cite{T1700296-v1,T1700297-v2,T1700298-v1,T1700299-v3} as well as O3A \cite{LIGO-T1900799-v1} and O3B \cite{LIGO-T2100286-v2}.

\begin{figure*} [htbp]
    \centering
    
    \includegraphics[width=\columnwidth]{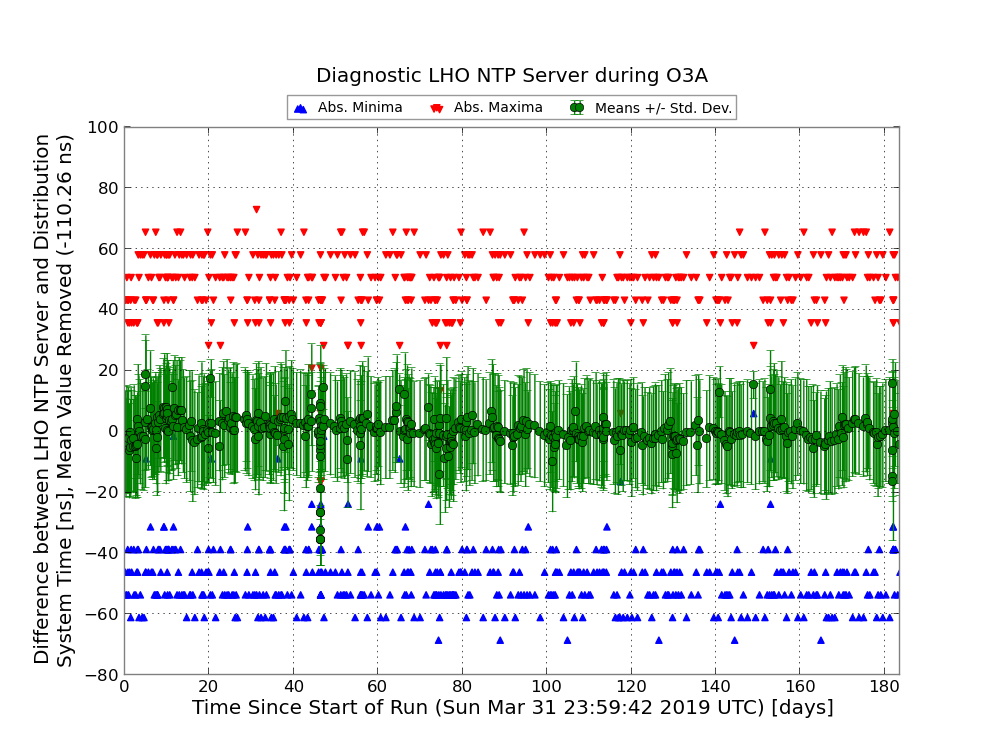}
    \includegraphics[width=\columnwidth]{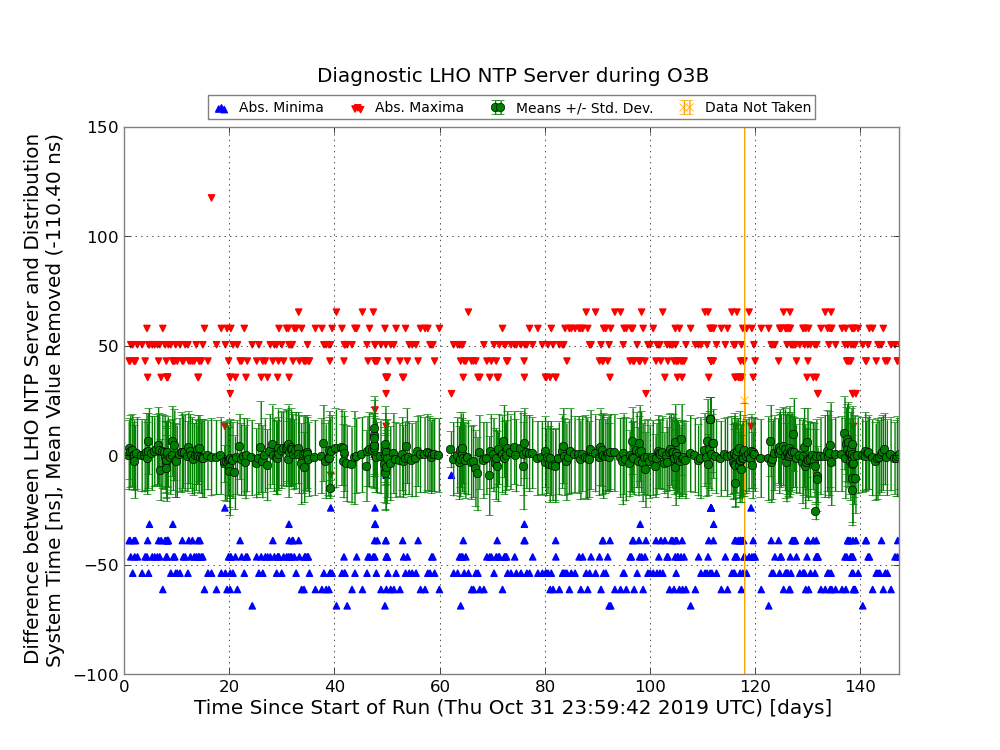}
    \includegraphics[width=\columnwidth]{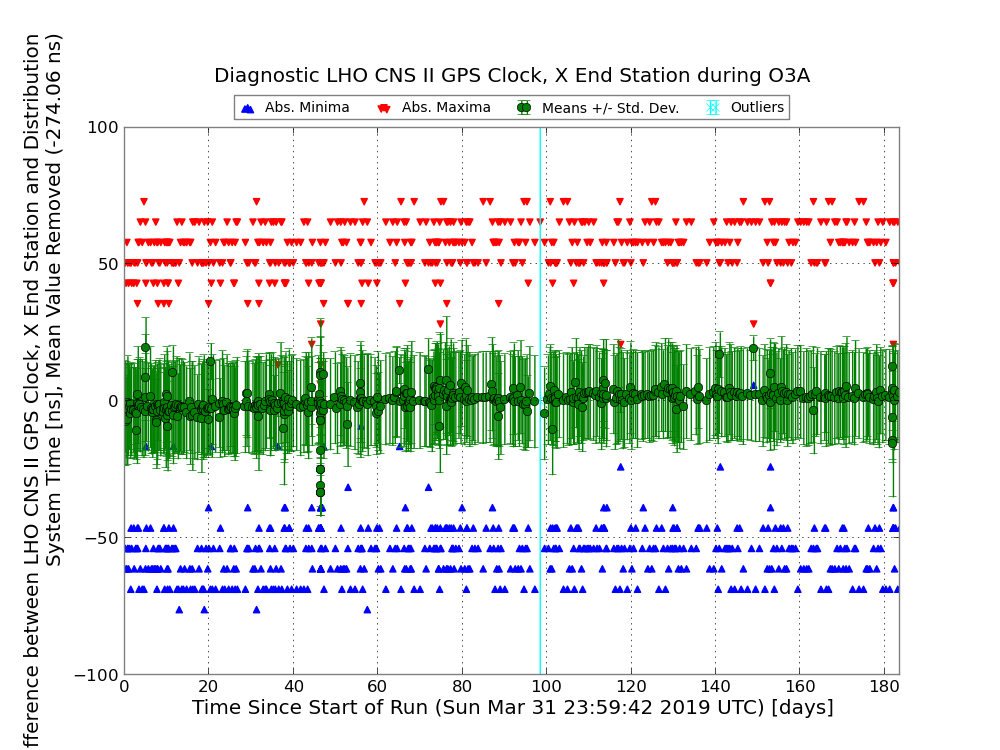}
    \includegraphics[width=\columnwidth]{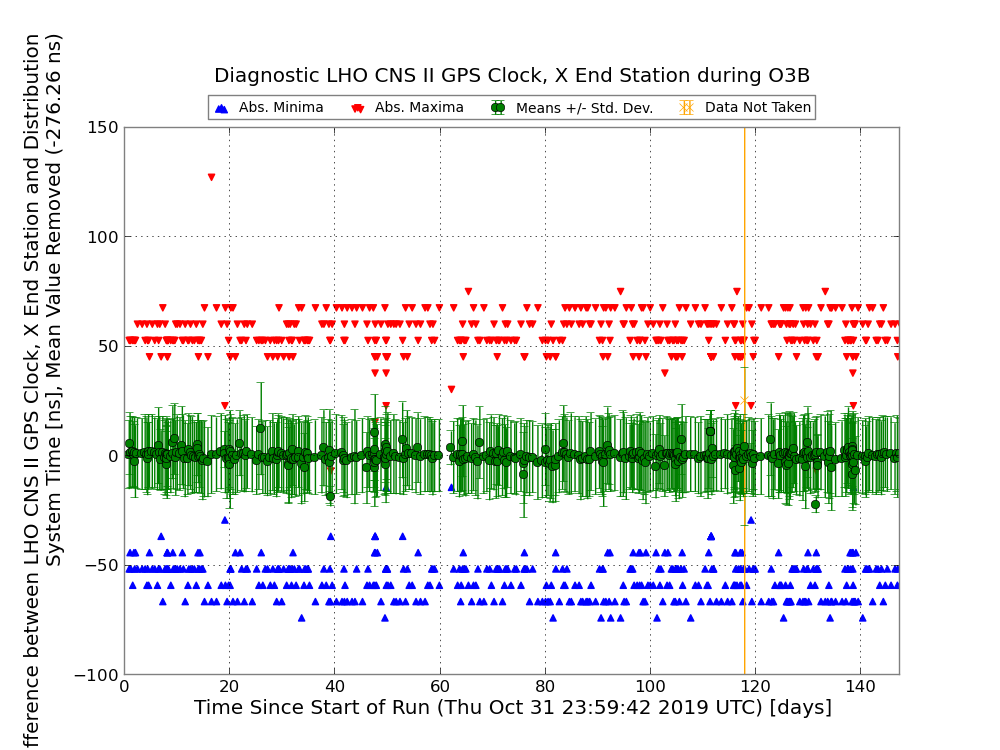}
    \includegraphics[width=\columnwidth]{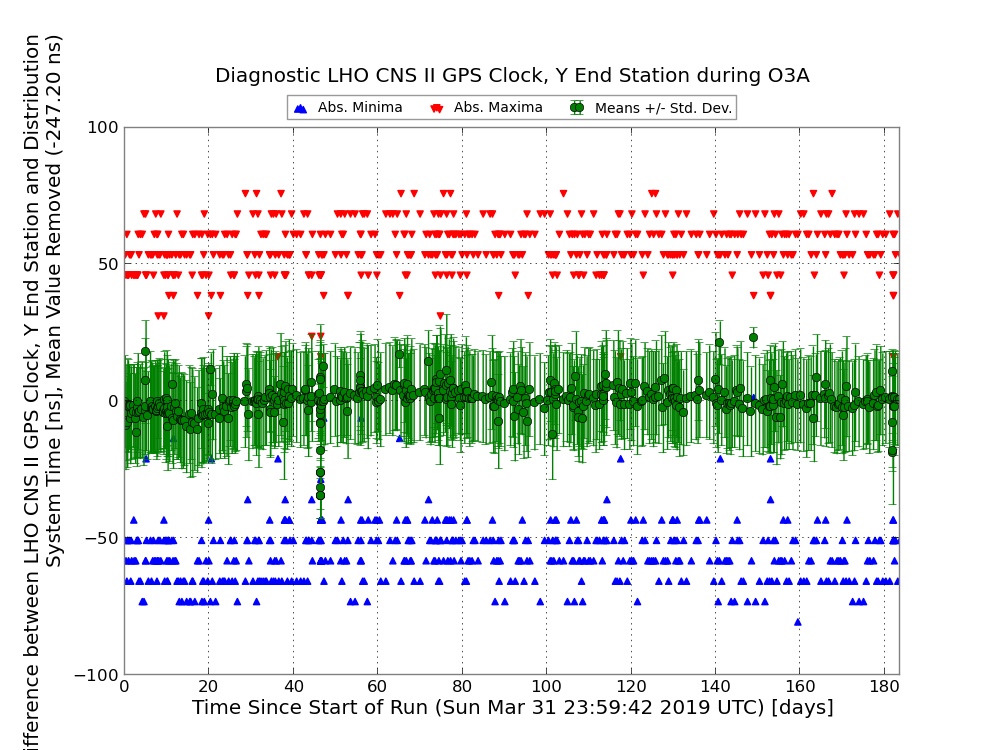}
    \includegraphics[width=\columnwidth]{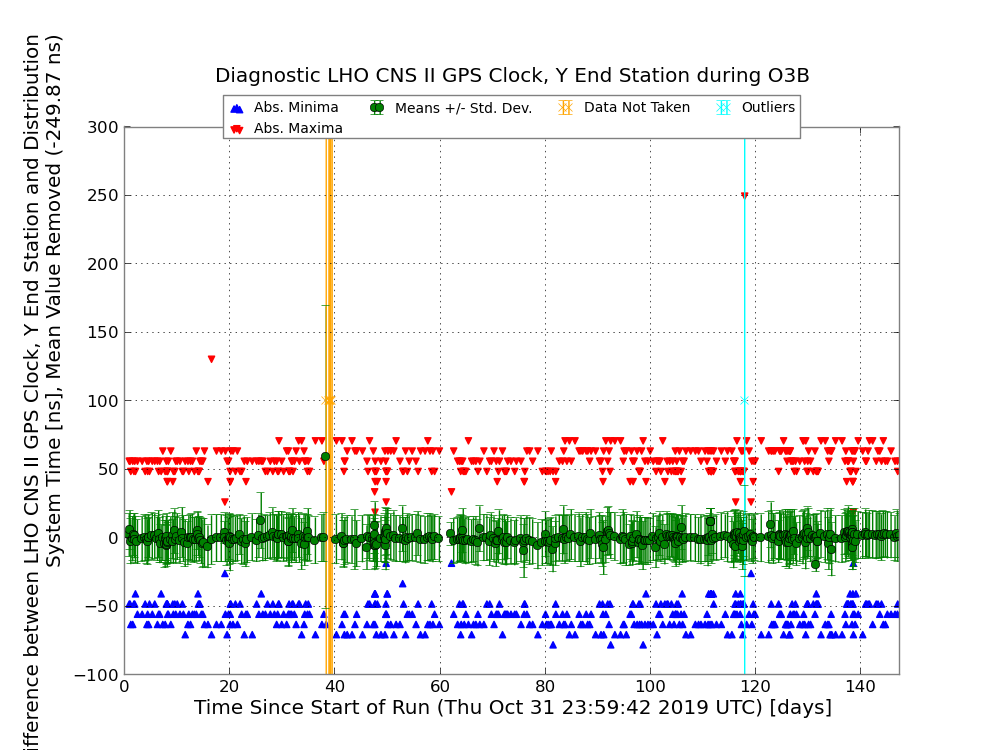}
    
    \caption{Plots of the timing difference between the LHO timing system 1PPS signal and various GPS clocks during O3, including (Top) time difference between the GPS-backed NTP Server and the timing system at the Corner Station, (Middle) time difference between a GPS-backed CNS II clock and the timing system at the X End Station, and (Bottom) time difference between a GPS-backed CNS II clock and the timing system at the Y End Station. O3A results are on the left and O3B results are on the right. The green dots represent the average time difference over the course of one lock segment, the error bars denote standard deviation of the average time difference, and the blue and red triangles denote the minimum and maximum time differences in each lock segment. The yellow line corresponds to a time when data was not taken in this channel and the light blue line corresponds to a time where the time difference exceeded the 1 $\mu$s threshold. The outlier at the X End Station 98 days into O3A was caused by a  GPS antenna issue where the antenna was briefly tracking 0 satellites and the outlier at the Y End Station 117 days into O3B was the result of a DAQ restart. The data not taken marker at 117 days of O3B on the NTP server and X End plots were due to the same DAQ restart. The data not taken marker 38 days into O3B on the Y End Station plot was caused by a failure in the fiber link between the slow controls Beckhoff chassis at the Y End Station.}
    \label{fig:LHO}
\end{figure*}

Figure \ref{fig:LHO} shows plots used to track the timing performance for LHO throughout O3. Figure \ref{fig:LHO} includes the time difference between the timing system and the CNS II GPS clocks at each End Station and the NTP Server. At the corner station and Y End Station during O3A, the timing variation was within $\pm$ 80 ns, well below the 1 $\mu$s performance threshold. At the X End Station during O3A all but one time segment showed variation below the 1 $\mu$s threshold. This outlier was due to a GPS antenna issue on July 8, 2019 when the antenna was momentarily tracking zero satellites and the CNS II clock returned faulty timing data. At the corner station and X End Station during O3B, the time difference was always within the 1 $\mu$s threshold. At only one time during the run did the time difference exceed the 1 $\mu$s threshold at the LHO Y End Station. This coincides with a DAQ restart on February 26, 2020. Besides this anomaly, the time difference between the GPS clocks and timing system at LHO was below $\pm150$ ns throughout O3B, with the overwhelming majority of time differences below $\pm80$ ns.

\begin{figure*} [htbp]
    \centering
    \includegraphics[width=\columnwidth]{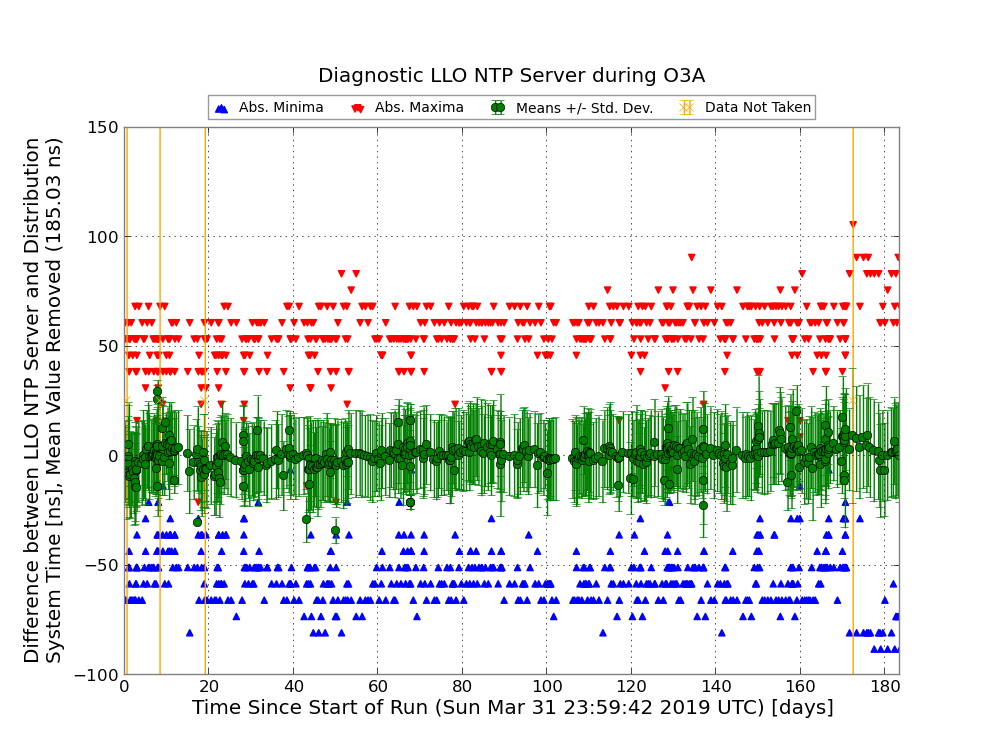}
    \includegraphics[width=\columnwidth]{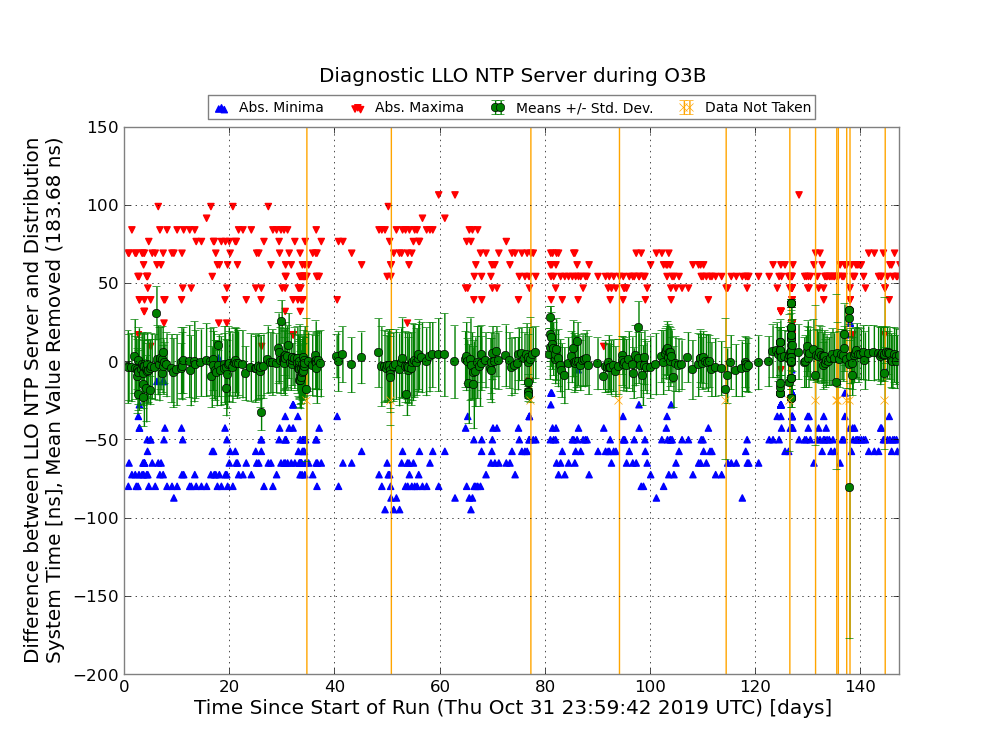}
     \includegraphics[width=\columnwidth]{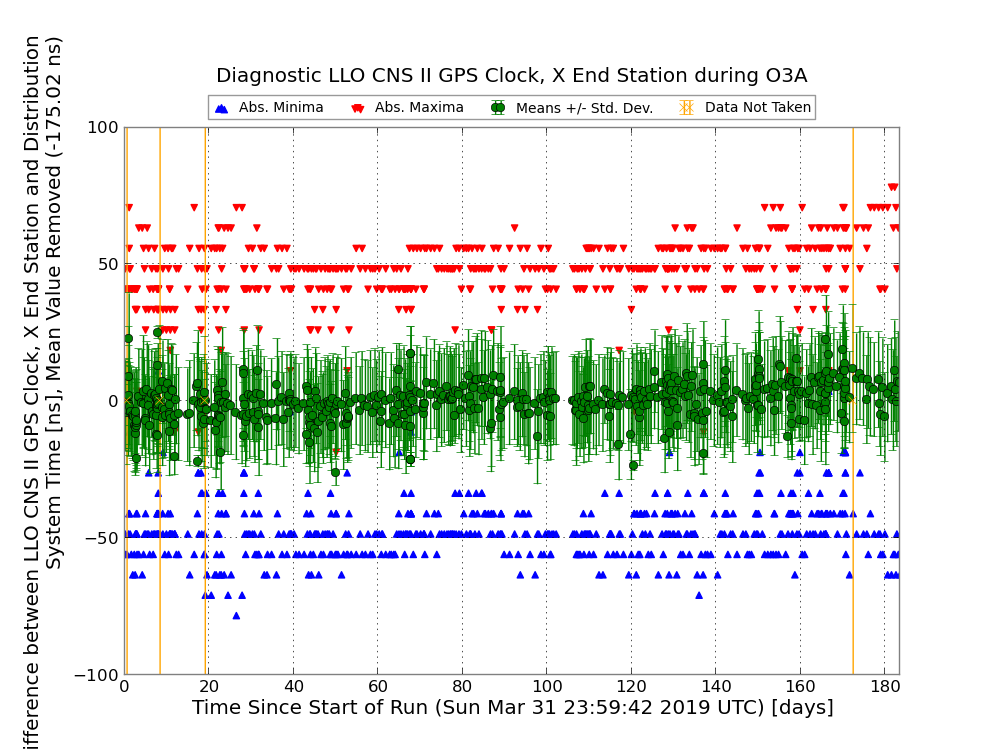}
    \includegraphics[width=\columnwidth]{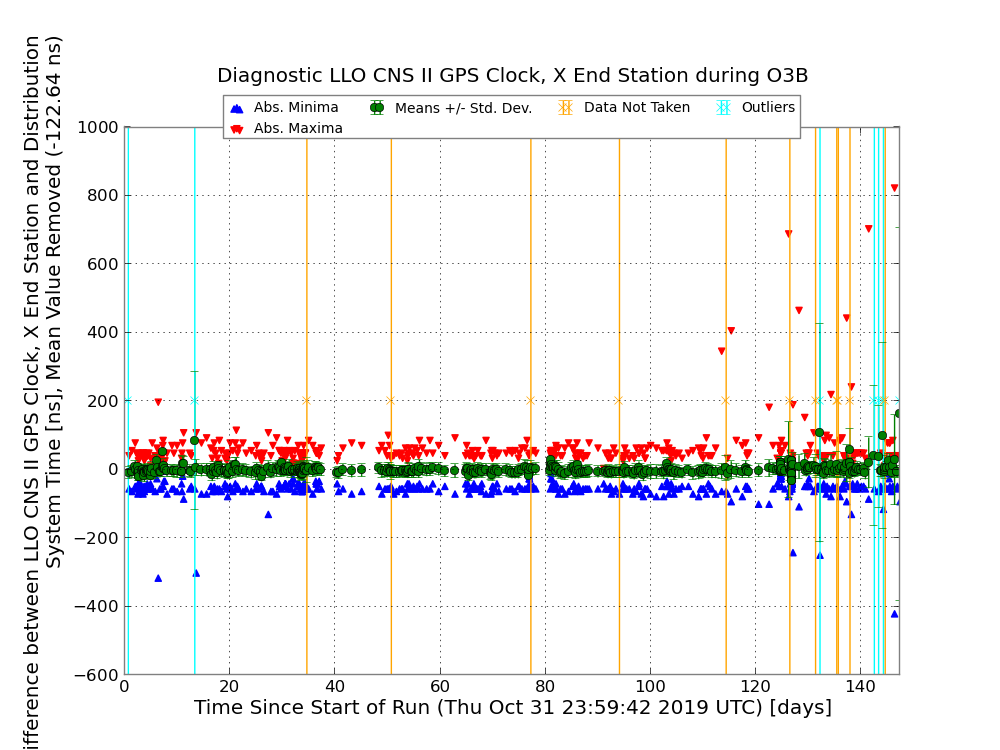}

    \caption{Plots of the timing difference between the LLO timing system 1PPS signal and various GPS clocks during O3, including (Top) time difference between the GPS-backed NTP Server and the timing system at the Corner Station and (Bottom) time difference between a GPS-backed CNS II clock and the timing system at the X End Station. O3A results are on the left and O3B results are on the right. The green dots represent the average time difference over the course of one lock segment, the error bars denote standard deviation of the average time difference, and the blue and red triangles denote the minimum and maximum time difference in each lock segment. The yellow line corresponds to a time when data was not taken in the channel and the light blue line corresponds to a time where the time difference exceeded the 1 $\mu$s threshold. All outliers in the X End Station plot were the result of GPS antenna issues as the antenna was tracking two or fewer satellites during those times. The data not taken lines in the O3A plots were caused by 1) a Beckhoff communication failure on April 1, 2019, 2) lock loss prior to preventative maintenance on April 8, 2019, 3) Beckhoff failure which caused lock loss on April 20, 2019, and 4) a DAQ process restart on September 20, 2019. The data not taken lines in the O3B plots were caused by 1) a DAQ process restart on December 5, 2019, 2) a Beckhoff connection issue which caused lock loss on December 21, 2019, 3) a DAQ restart on January 17, 2020, 4) a Beckhoff connection issue which caused lock loss on February 3, 2020, 5) a data concentrator failure and restart on February 23, 2020, 6) a DAQ screen meltdown and reset on March 6, 2020, 7) a DAQ screen meltdown and reset on March 11, 2020, 8) a DAQ restart on March 15, 2020, 9) a DAQ error on March 17, 2020, 10) a DAQ error on March 17, 2020 and 11) a DAQ restart on March 24, 2020.}
    \label{fig:LLO}
\end{figure*}

Figure \ref{fig:LLO} shows plots used to track the timing performance at LLO during O3. Figure \ref{fig:LLO} includes the time difference between the timing system and the CNS II GPS clock at the X End Station and the NTP Server. We omit the plots for the Y End Station because an issue whose origin is external to the timing system arose with the Y End redundant CNS II clock which caused repeated timing errors on the order of $10^{-1}$ s in every lock segment of O3. To confirm that the issue was with the CNS II GPS Clock and not the timing system, we ran DuoTone delay checks at times in both O3A and O3B where these persistent errors took place and found that the DuoTone delay returned the expected delays within the 1 $\mu$s threshold. 

At the LLO Corner Station and X End Station during O3A, all time variations were well within the $\pm$1 $\mu$s threshold. The Corner Station showed variation within $\pm$110 ns at all times and the X End Station showed variation within $\pm$80 ns at all times. At the Corner Station during O3B, the time difference was below the $\pm1$ $\mu$s threshold throughout the run and, for the vast majority of times, was below $\pm100$ ns. At the X End Station, there were 7 times in O3B during which the time difference exceeded the $1$ $\mu$s. These errors arose due to an issue with the X End Station GPS antenna connected to the CNS II clock and correspond to periods of time when the antenna was tracking between 2 and 0 satellites and providing unreliable GPS timing data. Additionally, a number of minima and maxima on the X End Station plot in Figure \ref{fig:LLO} exceeded $\pm200$ ns. The same GPS antenna issue caused the large scatter in the minima and maxima of lock segment time differences, as the GPS antenna was tracking two or fewer satellites at those times. None of these errors coincided with a GW event. Otherwise, for much of O3B, the time difference at the X End Station was within $\pm100$ ns. 

\begin{figure*} [htbp]
    \centering
    \includegraphics[width=\columnwidth]{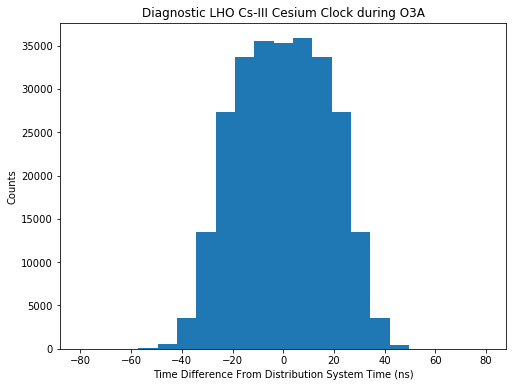}
    \includegraphics[width=\columnwidth]{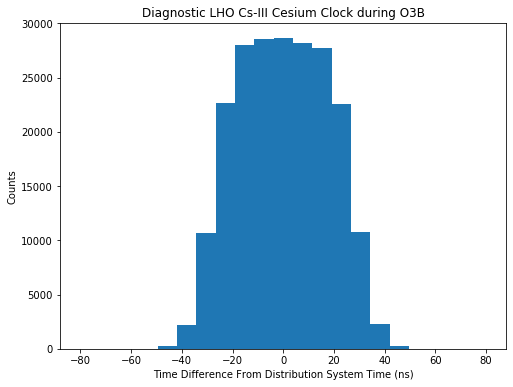}
    \includegraphics[width=\columnwidth]{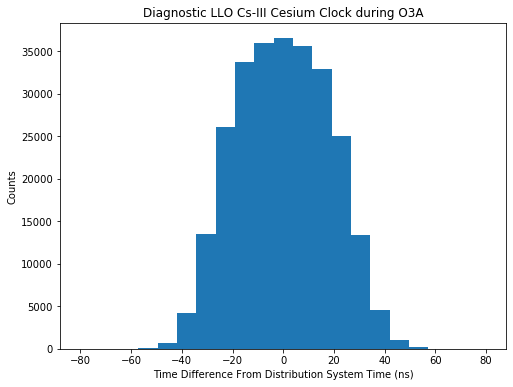}
    \includegraphics[width=\columnwidth]{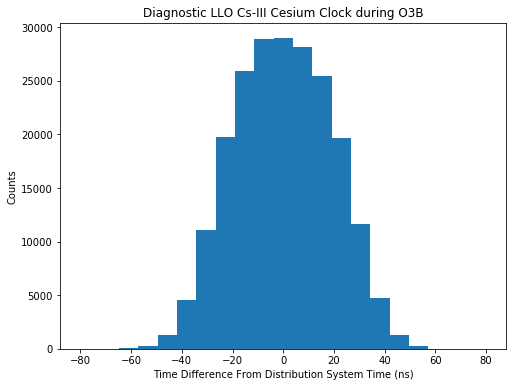}
    
    \caption{Histograms of the average time difference in every minute of O3 between the timing system and the time of the on-site Cs-III atomic clock with the drift trends removed in (Top) LHO and (Bottom) LLO. The left side shows O3A while the right shows O3B.}
    \label{fig:CsIII}
\end{figure*}

In addition to checking the timing system against GPS clocks at various stations, the timing at a given interferometer is verified against an on-site Cs-III atomic clock  \cite{csiiI4301B}. The Cs-III atomic clock is very precise on short time-scales; however, the Cs-III clock time drifts over the course of an observing run due to various environmental factors. In order to obtain accurate measurements of the deviation from the atomic clock time, the atomic drift trend is subtracted. Figure \ref{fig:CsIII} shows a histogram of the average timing system variation from the Cs-III time (with the trend removed) for each minute of O3 (excluding minutes when timing data was unrecorded) at LLO and LHO. The timing variation between the atomic clock and timing system was always within $\pm$80 ns  at both LHO and LLO during O3.

% -------------------------------------------
\section{Conclusion}
\label{sec:DiscussionandConc}
% -------------------------------------------
Throughout O3, the LIGO timing system displayed excellent timing precision. The DuoTone delay and IRIG-B checks confirmed that the timing system contributed less than $\pm$1 $\mu$s to the total error in the GW arrival times during observed GW events. Consequently, the astrophysical results for the O3 GW events are not limited by timing precision. 

The long term performance of the timing system was well within the designed specifications. The overwhelming majority of times throughout both O3A and O3B displayed timing errors of less than $\pm$1 $\mu$s at both LLO and LHO, and in most cases the system performed much better. General control system and GPS antenna issues affected timing system data during the run, but these problematic times were diagnosed. None coincided with GW events as they were confined to a minuscule fraction of the observing run. Most importantly, these issues were external to the timing system. 

In fact, the precision and consistency of the timing system allowed us to diagnose issues with other devices such as GPS antennae and external GPS clocks. In addition to its excellent time-keeping ability, the timing system serves as an investigative tool to study errors in other parts of the instrument. 

Through three complete observing runs, the LIGO timing distribution has successfully provided critical information for the astrophysical discoveries of LIGO. The timing system met and even exceeded mission requirements. This precise timing information provided to the interferometers enabled the first recorded detections of GW transients and will continue to allow for GW detections in upcoming observing runs.

% -------------------------------------------
\section*{Acknowledgments}
% -------------------------------------------
This material is based upon work and data supported by NSF's LIGO Laboratory which is a major facility fully funded by the National Science Foundation. LIGO is funded by the U.S. National Science Foundation. Virgo is funded by the French Centre National de Recherche Scientifique (CNRS), the Italian Istituto Nazionale della Fisica Nucleare (INFN) and the Dutch Nikhef, with contributions by Polish and Hungarian institutes.

The authors are grateful for the LIGO Scientific Collaboration review of the paper and this paper is assigned a LIGO DCC number(LIGO-P2200388). 

Special thanks to Evan Goetz and Shivaraj Kandhasamy for their help.

The authors are grateful to the Operation Specialists of the LIGO Hanford and Livingston observatories for their extraordinary efforts that make observations possible. The authors would like to thank Jonathan Hanks, Patrick Thomas, Corey Gray, Stuart Aston, and Erik R. G. von Reis for their contributions.

The authors acknowledge the LIGO Collaboration for the production of data used in this study and the LIGO Laboratory for data generation on its computing resources (National Science Foundation Grants PHY-0757058 and PHY-0823459). The authors are grateful to the LIGO DAQ, Commissioning and Detector Characterization Teams and LSC domain expert Colleagues whose fundamental work on the LIGO detectors enabled the data used in this paper. The authors would like to thank colleagues of the LIGO Scientific Collaboration and the Virgo Collaboration for their help and useful comments with special emphasis on Colleagues at the LIGO Hanford and Livingston Observatories. The authors would like to acknowledge the critical contributions Paul Schwinberg of LHO.

The authors thank Columbia University in the City of New York for their generous support.
The Columbia Experimental Gravity group is grateful for the generous support of the National Science Foundation under grant PHY-1708028. A.S. is grateful for the support of the Columbia College Science Research Fellows program and the Heinrich, CC Summer Research Fellowship. Y.A. acknowledges the support of a Grant-In-Aid of Research from Sigma Xi, The Scientific Research Society. I.B. acknowledges the support of the Alfred P. Sloan Foundation and NSF grants PHY-1911796, PHY-2110060, and PHY-2207661.

We would like to thank all of the essential workers who put their health at risk during the COVID-19 pandemic, without whom we would not have been able to complete this work.

% -------------------------------------------
\bibliography{References,LIGO_PP_Detector_References} 
% -------------------------------------------

\end{document}